\documentclass{aa}
\usepackage{graphicx}
\begin{document}
\title{SN Ib 1990I:  Clumping and Dust in the Ejecta? }
\titlerunning{SN 1990I }  

 \author{A. Elmhamdi \inst{1,2}, I.J. Danziger \inst{3}, E. Cappellaro \inst{4}, M. Della Valle \inst{5}, C. Gouiffes \inst{6},\\ Mark M. Phillips \inst{7}, M. Turatto \inst{8}}

   \offprints{A. Elmhamdi, {elmhamdi@sissa.it}}
   
   \institute{SISSA / ISAS , via Beirut 4 - 34014 Trieste - Italy
     \and European Southern Observatory, Karl-Schwarzschild-Str 2, 85748 Garching b. Munchen, Germany
      \and INAF, Osservatorio Astronomico di Trieste, Via G.B.Tiepolo 11 - I-34131 Trieste - Italy 
       \and INAF, Osservatorio Astronomico di Capodimonte, via Moiariello 16,
         I-80131 Napoli, Italy
      \and INAF, Osservatorio Astrofisico di Arcetri, largo E. Fermi 5, 50125
	   Firenze, Italy   		
      \and CEA/DSM/DAPNIA/Service d'Astrophysique, Saclay, Paris, France 	     \and Las Campanas Observatory, Carnegie Observatories, Casilla 601, La serena, Chile
      \and INAF, Osservatorio Astronomico di Padova, vicolo dell'Osservatorio 5,
         35122 Padova, Italy	
}
 
\authorrunning{Elmhamdi et al.}

\date{}

\abstract{ Photometry and spectra of the type Ib SN 1990I are presented and analysed, covering about 400 days of evolution. The good quality of the data allow one to set reliably the supernova age (i.e. both date of maximum and explosion time estimates are given). 
The presence of optical helium lines is shown. SN 1990I seems to show higher velocities compared to a sample of type Ib events. The nebular emission lines display a high degree of asymmetry and the presence of fine structures, suggestive of non-spherical clumping in the ejecta of SN 1990I. Using the [O I] 6300,64 \AA~flux, we estimate a lower limit on the oxygen mass to fall in the range $(0.7-1.35)$ M$_\odot$. The oxygen mass requires a filling factor as small as $\sim$10$^{-2}$ on day 254, indicating a highly clumpy distribution of the oxygen material. The amplitude and evolution of the [Ca II]/[O I] flux ratio is similar to that observed in a sample of SNe Ib/c events and thus  suggests similar progenitor masses. A blueshift of the order 600 km s$^{-1}$ is reported in the [O I] 6300,64 \AA~after day 254. The [Ca II] 7307.5 \AA~emission profile appears blueshifted as well at late epochs. We recover the quasi-bolometric ``$BVRI$'' light curve of SN 1990I. The constructed bolometric light curve shows a change of slope at late phases, with an $e$-folding time of 60 $\pm$2 d in the $[50:200]$ d time interval, considerably faster than the one of $^{56}$Co decay (i.e. 111.3 d), suggesting the $\gamma$-rays escape with lower deposition, owing to the low mass nature of the ejecta. After day 200, an $e$-folding time $\simeq$47 $\pm$2.8 d
is measured. While the light curves of SNe 1990I and 1993J are similar in the $[30:100]$ d time range, they tend to behave differently after day 200. A simplified $\gamma$-ray deposition model is applied after adding a contribution of about 35$\%$ to the computed pseudo-bolometric light curves to account for near-IR luminosities to estimate the ejecta and $^{56}$Ni masses ($M(^{56}Ni)=0.11 ~$M$_\odot$ and $M_{ej}=3.7~ $M$_\odot$). The deficit in luminosity is estimated to be about 50$\%$ around day 308.
The observed spectral blueshift combined with the dramatic and sudden drop in the pseudo-bolometric light curve and $(B-V)$ colour is interpreted to be a consequence of dust condensation in the ejecta of SN 1990I around day 250.  

}
\maketitle
\section{Introduction}
  One of the interesting topics in supernovae (SNe) research for the last few years has been the study of hydrogen deficient events, namely type Ib (helium-rich) and type Ic (helium-poor) events (Wheeler $\&$ Harkness. 1990)\nocite{Whel90}.
The particular interest in these two subclasses comes from the fact that they are the least understood among supernovae varieties in their observational properties, progenitors and therefore the physics of the explosion (Nomoto et al. 1990\nocite{Nom90}; Baron. 1992\nocite{Bar92}). The possible connection of the strongly energetic ones with Gamma-Ray Bursts (i.e. the emergence of Hypernovae; Woosley et al. 1999\nocite{Woos99}; Nomoto et al. 2000\nocite{Nom00}) makes them exciting events. In addition, the appearance of the transition events such SNe 1993J and 1987K suggests a physical connection between type Ib/c and type II objects; thus one needs to understand the differences as well as the similarities among them.

Type Ib and Ic SNe are classified on the basis of their spectra. These events are hydrogen deficient at maximum light. They lack also the deep Si II absorption near 6150 \AA~ characterizing type Ia events. SNe Ic are spectroscopically similar to SNe Ib at late epochs where the spectra are dominated by oxygen and calcium emission lines, whereas at early phases SNe Ic do not show the He I lines (Wheeler $\&$ Harkness. 1990).

Different models, evolution and explosion, have been suggested to explain the main observational properties of SNe Ib/c. The most accepted scenario is related to the mechanism of the hydrogen rich SNe II, namely core collapse in massive stars. There are many arguments reinforcing the idea of core collapse events as the favoured in the progenitors of type Ib/c SNe :
\begin{figure*}
\includegraphics[height=10.5cm,width=9cm]{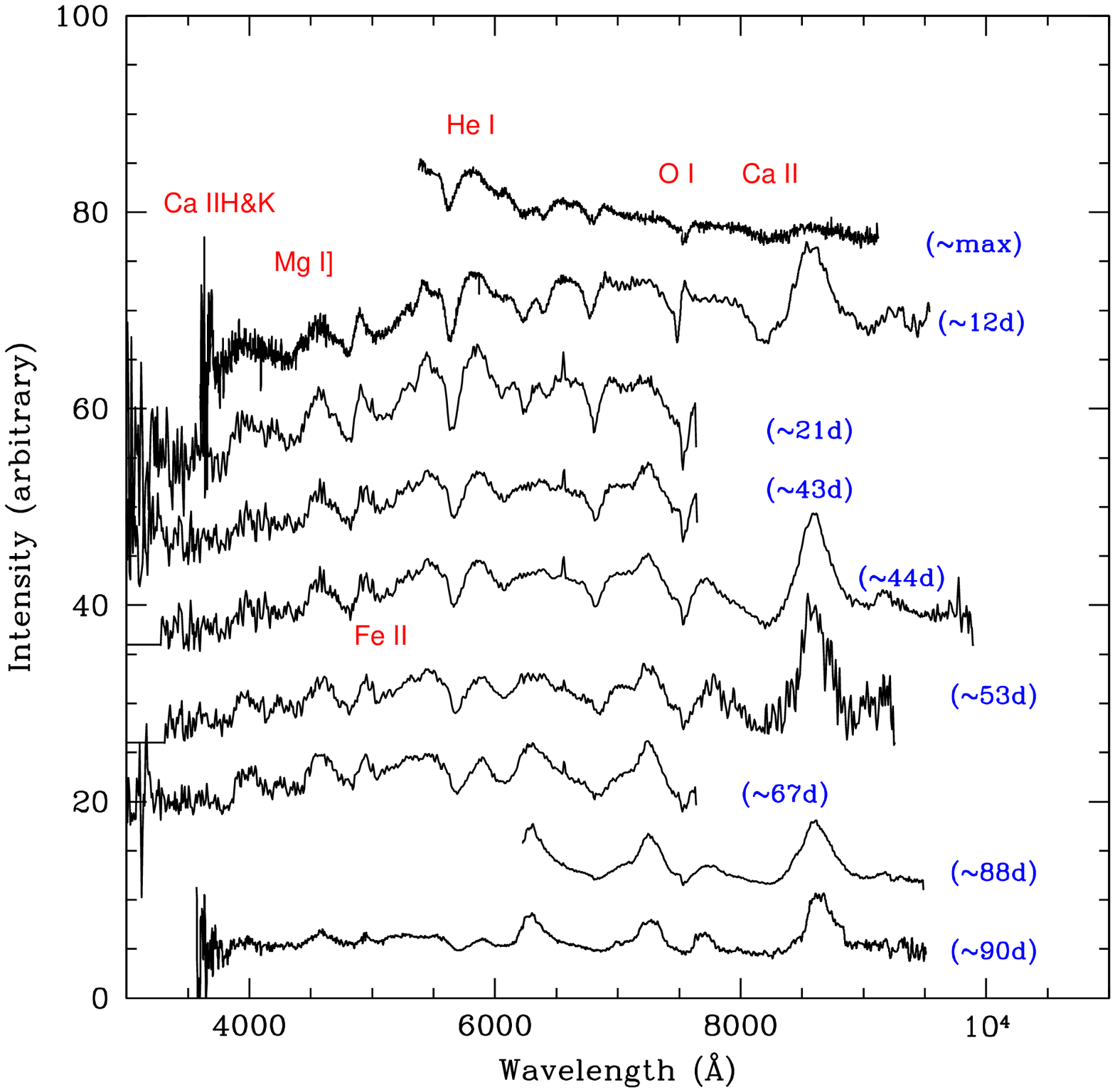}\includegraphics[height=10.5cm,width=9cm]{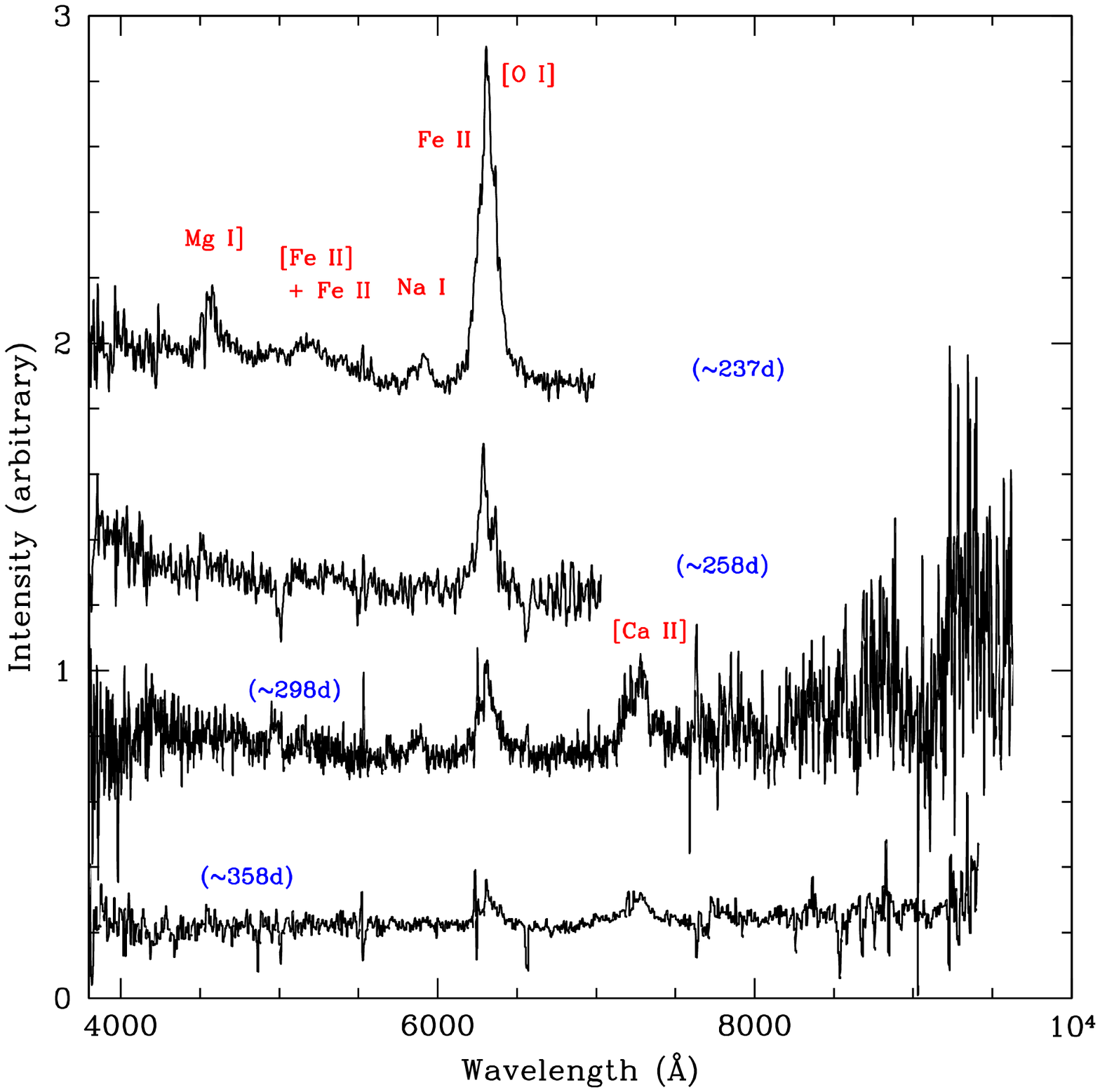}
\caption{ Spectral evolution of SN 1990I during the photospheric phase (A), and during the nebular phase (B). Days since maximum are shown. All the spectra have been corrected for the recession velocity of the host galaxy and scaled arbitrary for clarity. The strongest features are also reported.}
\end{figure*}                                      

$-$ They are associated with population I stars, and are observed to occur in spiral galaxies ongoing star formation occurs (association with H II regions).

$-$ The presence of strong oxygen emission lines. 

$-$ The discovery of transition events (e.g. SNe 1987K and 1993J; Filippenko 1988\nocite{Fil88}, Swartz et al. 1993\nocite{Swar93}), whose spectral properties changed from type II to type Ib/c as they aged.

$-$ The early detection of such events in the radio (e.g. SN 1983N and SN 1984L), suggests interaction with a dense circumstellar material, hence strong mass loss. 

Here we present and analyse the photometry and spectral evolution of type Ib SN 1990I. The paper is organized as follows. We describe first the spectral behaviour of SN 1990I during both the photospheric and nebular phases. We use the [O I] 6300,64 \AA~luminosities to estimate the ejected mass of oxygen (Sect. 2.1.2). Then we study the light curves and colour evolution of SN 1990I. The photometry is compared to a selected sample of type Ib/c, IIb events (Sect. 2.2). We finally describe a simple model of $\gamma$-ray deposition from radioactive decay used to fit the bolometric evolution of SN 1990I, providing $^{56}$Ni and ejecta mass estimates (Sect. 3.2). 
\section{Observations}

SN 1990I was discovered by Pizarro et al. (IAUC 5003)\nocite{Piz50} in NGC 4650A. Initially it was classified as type Ia on the basis of an early spectrum taken by Sivan and Marseille on April 30 (IAUC 5003). Based on three early spectra (May 18, June 7 and June 8) taken with the Cerro Tololo 1.5-m telescope, SN 1990I has been re-classified as a type Ib by Phillips (IAUC 5032)\nocite{Phil50}. The SN is located in the polar ring of its host ring galaxy NGC 4650A. 

We present spectra and photometry covering about $\sim$ 400 d of evolution. The data are collected both from ESO-KP (Key Project) database and from CTIO observations. 
\subsection{Spectroscopic evolution}
\begin{table} \caption{Spectroscopic observations of SN 1990I}
\begin{tabular}{c c c c }
\hline \hline
 Date (dd/mm/yy) & Phase$^{\dagger}$   &  Range (\AA)& Instrument\\
\hline 
29/04/90   & $\sim$ max& 5370-9100  & ESO-2.2m \\
09/05/90   & $+$ 12d & 3610-9510  & ESO-NTT \\
18/05/90   & $+$ 21d& 2950-7620  & CTIO-1.5m \\
07/06/90   & $+$ 43d&  2950-7600 & CTIO-1.5m \\
08/06/90   & $+$ 44d&  3310-9840 & CTIO-1.5m \\
19/06/90    & $+$ 53d&  3320-9210 & CTIO-1.5m \\
02/07/90    & $+$ 67d&  2960-7620 & CTIO-1.5m \\
24/07/90    & $+$ 88d&  2960-7620 & CTIO-4.0m \\
26/07/90    & $+$ 90d&  6230-9460 & CTIO-4.0m \\
21/12/90   & $+$ 237d&  3580-6940 & ESO-3.6m\\
11/01/91   & $+$ 258d&  3590-7010 & ESO-3.6m\\
20/02/91   & $+$ 298d&  3700-9590 & ESO-3.6m\\
21/04/91   & $+$ 358d&  3530-9390 & ESO-3.6m\\

\hline
\hline
\end{tabular} \\[1ex]
\footnotesize
\emph{\rm $\dagger$ since the maximum light time: $\sim$ 29/04/1990}\\ 

\normalsize
\end{table}

The journal of the spectroscopic observations  of SN 1990I are reported in Table 1. Figure 1  illustrates the spectroscopic evolution, covering both the photospheric and nebular phases.  The well sampled data of SN 1990I, especially light curves, allow us to estimate the maximum light time with some confidence (see Sect. 2.2 for discussion regarding the SN age). For each spectrum the corresponding days since maximum light are indeed reported . The spectra have been corrected for the recession velocity ($cz=$ 2902 km s$^{-1}$) of the parent galaxy NGC 4650A.

The early spectrum (around maximum) displays a blue continuum with broad P-Cygni profiles, indicating both high velocities and temperature. The He I 5876 \AA~ is unambiguously recognized with a velocity of the order $\sim$ 13400 km s$^{-1}$, as measured from the absorption minimum, clearly high compared to other SNe Ib at similar phase ($\sim$ 10000 km s$^{-1}$; Branch et al. 2002, hereafter B02)\nocite{Bra02}. For the 12 d spectrum, V$\sim$ 12500 km s$^{-1}$ is measured, and this is used as reference to infer the possible location of the various He I lines, namely 4471 \AA, 5015 \AA, 6678 \AA~ and 7065 \AA. Figure 2 demonstrates the evidence of the prominent optical He I lines in the ejecta of SN 1990I. The relative strengths of the He I lines are such that strong departures from LTE are apparent as suggested by Wheeler et al. (1987)\nocite{Whee87} for other type Ib SNe. Lucy (1991)\nocite{Lucy91} has shown that strong departures from LTE result from excitation  by non-thermal electrons
originating from  radioactive decay of $^{56}$Co dispersed throughout the
envelope. It must be kept in mind, nevertheless, that since the Fe II lines, especially 4924 \AA, 5018 \AA, affect mainly the spectra shortward 5200 \AA, the absorption features are subject to line blending in that portion of the spectra. In addition the He I 4471 \AA ~line may be affected by the Mg II 4481 \AA.     
\begin{figure}
\includegraphics[height=8.5cm,width=8cm]{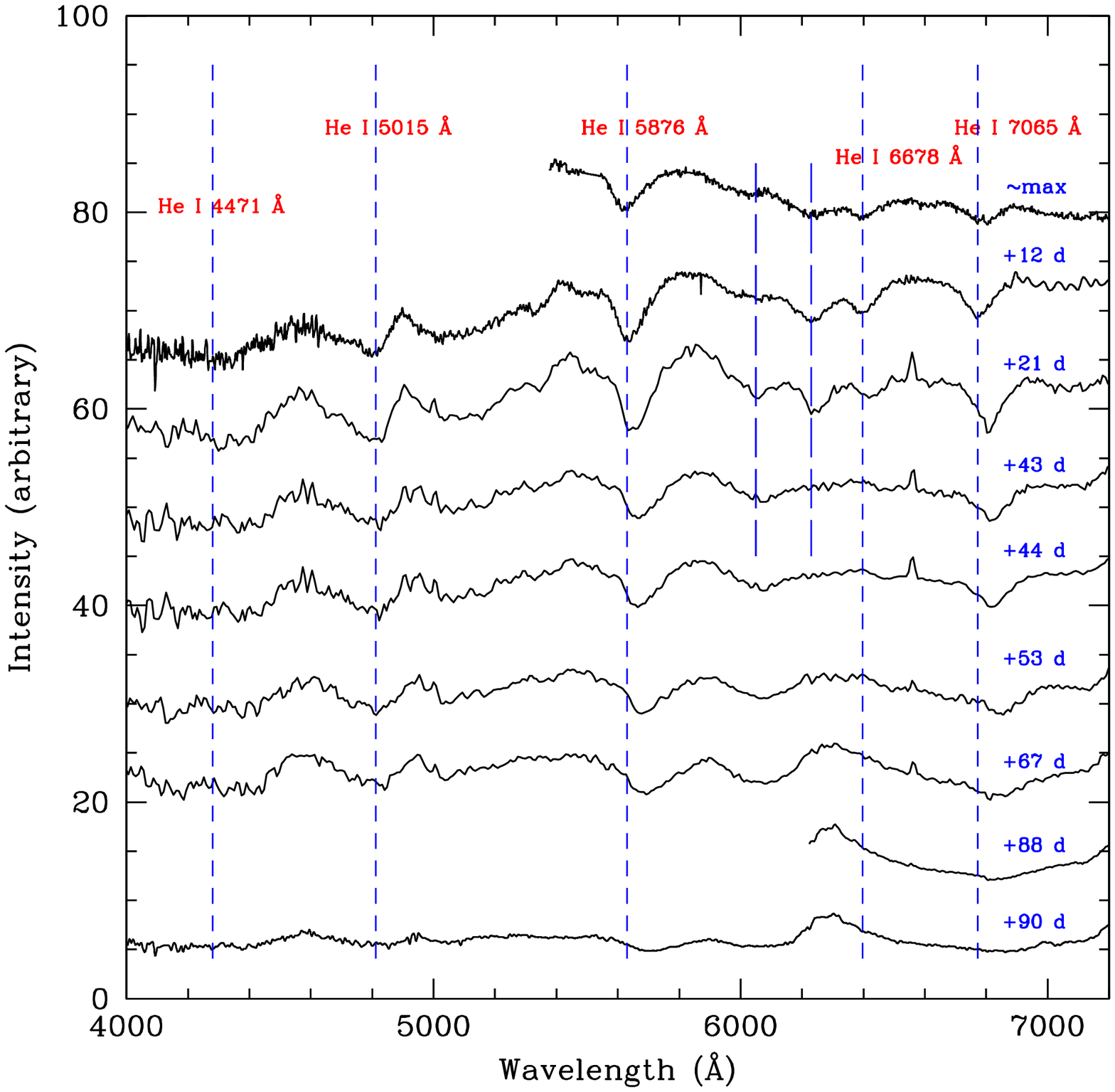}
\caption{ Possible location of He I lines in the photospheric phase of SN 1990I (short-dashed lines). The absorption features near 6050 \AA~ and 6250 \AA~ are marked (long-dashed lines; Section 2.1). Phases since maximum light are reported. }
\end{figure}

The photospheric velocity, as inferred from Fe II 5169 \AA ~line, appears to be greater with respect to values derived from the SNe Ib sample in the B02 work. From the 12 d spectrum we measured V$_{phot}\sim$ 9000 km s$^{-1}$. At comparable phase B02 provided V$_{phot}\sim$ 6500 km s$^{-1}$. If we use the derived velocity as an indicator of the epoch, and hence try to fit the B02 sample data (their Figures 22 and 23), one estimates the epoch to be around maximum. This seems unlikely, taking into account both the nature of the spectrum and the shape of the light curves (Sect. 2.2). Therefore the high velocity nature of SN 1990I ejecta seems unavoidable.  

The Ca II near-IR triplet, displaying broad P-Cygni profiles, is the strongest emission feature at red wavelengths at this phase. The Ca II H $\&$ K is also present. Although features in the blue part of the spectrum are hardly discernible because of the noise, we remark in the 21 d spectrum a weak absorption feature centered at $\sim$ 3890 \AA. This feature has been a subject of discussion for early spectra of SN Ib 1999dn, and has been attributed to Ni II 4067 \AA ~in order to fit the 3930 \AA ~absorption (Deng et al. 2000; their Sep. 14 spectrum)\nocite{Den00}. This best fit at 3930 \AA~with Ni II 4067 \AA~, for the case of SN 1999dn, leads the authors to believe it is early evidence of $^{56}$Ni mixing. 

In Figure 2 two absorption features near 6050 \AA ~and 6250 \AA ~are marked (long-dashed lines). Both these features are present in the three early spectra, while the 6250 \AA~feature is no more seen in the 43 d spectrum. Deng et al. (2000), through spectrum synthesis of SN 1999dn around maximum, have argued for evidence of both C II and H$\alpha$, attributing the feature around 6300 \AA ~to the blend of the two lines, and uniquely due to C II later on. Benetti et al. (2002)\nocite{Ben02}, on the other hand, have included Ne I lines to fit early spectra of SN 1991D, providing improvement in fitting both absorption features, similar to the ones marked by long dashed lines in Fig. 2. 

Whether hydrogen is present or not in type Ib/c events is an important issue, since it is directly related to the nature of the progenitor stars and their evolution (Filippenko 1997)\nocite{Fil97}. 

The possible identification of the feature around 6250 \AA ~as H$\alpha$ may point to the presence of a thin hydrogen layer above the helium layer with high velocity, which disappears or is overwhelmed by C II 6580 \AA ~as H$\alpha$ optical depth decreases as a consequence of the envelope expansion (Deng et al. 2000). In fact, if correctly identified, we measure from the 12 d spectrum an H$\alpha$ ~velocity of $\sim$ 15000 km s$^{-1}$, higher compared to that of He I 5876 \AA ($\sim$ 12500 km s$^{-1}$). Note that since SN 1990I is very similar to SN 1991D with respect to the presence and the shape of the two absorption features one might expect Ne I lines to provide good fit (Benetti et al. 2002), while the presence of H$\alpha$ still remains a possible interpretation. Detailed synthetic spectra are needed to test both possibilities.      
  
The absorption near 7500 \AA ~is attributed to O I 7774 \AA ~line in type Ib SNe, nevertheless in some cases the introduction of the blend with Mg II 7890 \AA ~is needed for fitting using synthetic spectra (e.g. SN 1999dn, Deng et al. 2000). 

As the supernova evolves in time the features display a shift towards the red part of the spectrum. This shift is caused  by the retreat of the photosphere into the slower moving layers.

The last three spectra in Figure 1(left panel) present evidence of [O I] 6300,64 \AA ~and [Ca II] 7291,7324 \AA ~emissions, suggesting the onset of the nebular phase. The right panel, on the other hand, displays the real nebular phase spectra. As illustrated, the 237 d spectrum, displays a strong [O I] 6300,64 \AA ~emission line, which is quite broad.

 The other prominent features are Na I D 5893 \AA ~doublet and the broad Mg I] 4571 \AA~ line. Additional nebular lines are difficult to discern.
Regrettably, the 237 d spectrum does not extend redward of 7000 \AA, whereas the 298 d spectrum does. In fact, although the spectra are noisy, the [O I], [Ca II], Na I and Mg I] emission lines are still stronger compared to the continuum, and as time goes on they tend to shrink with respect to the continuum. This continuum may come from stars very close to the SN. In fact the presence of narrow H$\alpha$ emission in at least five photospheric and either H$\alpha$ emission or absorption in three of the nebular spectra suggests the presence of an H II region inadequately subtracted. The narrow H$\alpha$ feature ($EW \simeq 3.8$ \AA~on day 21) is located at the rest wavelength of the host galaxy and is seen to not broaden with time, which reflects its association with H II region rather than with the SN. 

Interestingly, at this phase, the [Ca II] and [O I] fluxes seem to be comparable. This is of interest since it is related to the nature of the progenitor star. Fransson $\&$ Chevalier (1987, 1989)\nocite{Fra87} have shown that the forbidden emission line ratio [Ca II]/[O I] is weakly dependent on density and temperature of the emitting zone, and is expected to show an almost constant evolution at late epochs. Furthermore the ratio seems to be very sensitive to the core mass and hence to the main-sequence one. The ratio increases with decreasing main-sequence mass. In Figure 3 we show the evolution of the computed ratio in SN 1990I together with SNe 1985F(Ib), 1998bw(Ic), 1993J(IIb), 1996N(Ib) and 1987A(II). The plot is indicative of the similar main-sequence-mass-nature of SN 1990I compared to the other Ib/c events. 

It is note worthy, however, that in type Ib/c SNe there is no hydrogen rich Ca II emitting zone as is the case of type II events (de Kool et al. 1998)\nocite{dek98}. We note as well that the values reported for SN 1990I may be even lower if we correct for the depletion behaviour blueward of the [O I] emission (see Section 2.1.1).
 
\begin{figure}
\includegraphics[height=8cm,width=8cm]{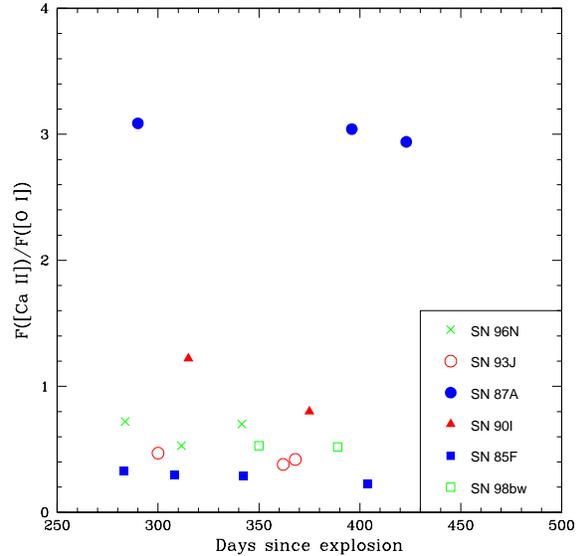}
\caption{The time evolution of the ratio $F([Ca II])/F([O I])$ of SN 1990I together with that of a sample of SNe. The difference between SN 1987A and the remaining SNe is clear. For convenience we refer to phases since explosion date (see Sect. 2.2.).} 
\end{figure}

\subsubsection{Nebular emission line profiles}
In Figure 4 we show, in velocity space, the late evolution of the most prominent nebular lines, namely Mg I] 4571 \AA, [O I] 6300,64 \AA, [Ca II] 7307.5 \AA ~and O I 7774 \AA. The [O II] 7322 \AA~, on the other hand, may contribute in the red side of the [Ca II] 7307.5 \AA.
\begin{figure*}
\center
\includegraphics[height=13.5cm,width=15.5cm]{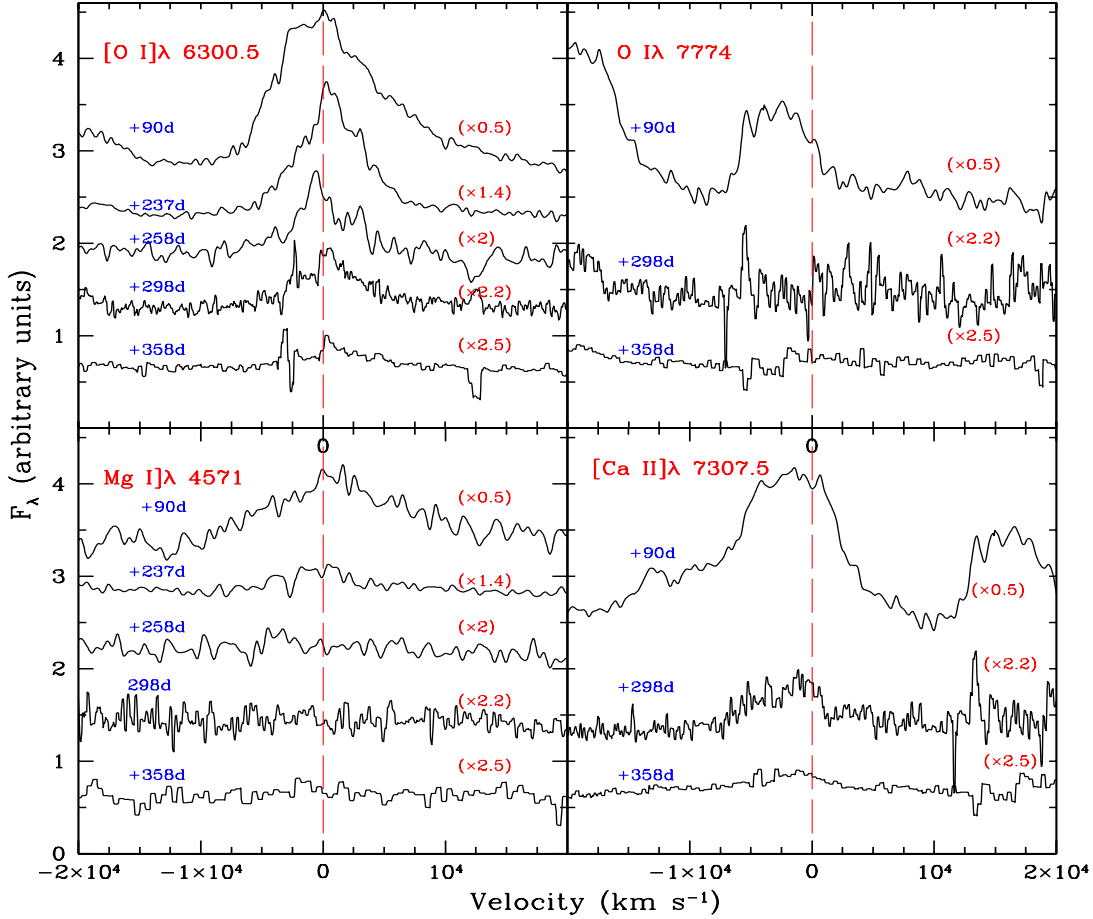}
\caption{ The time evolution, in velocity space, of the most prominent nebular lines: Mg I] 4571 \AA, [O I] 6300,64 \AA, [Ca II] 7307.5 \AA ~and O I 7774 \AA.The [O II] 7322 \AA~may contribute in the red side of the [Ca II] 7307.5 \AA. Shown as well are the coefficient by which the spectra have been multiplied for clarity. Rest wavelengths are marked by vertical long-dashed lines.} 
\end{figure*}

At 90 d the flat top of the [O I], [Ca II] and O I emission lines is indicative of shell structure. An approximate position of the blue edge of plateau in [O I] line is at 2600 km s$^{-1}$. The [O I] line on day 237 is still peaked at zero velocity, whereas on day 258 it appears blueshifted by $\sim$ 600 km s$^{-1}$. Unfortunately, the spectra in this transition phase, from +237 d to +258 d, do not show [Ca II] and O I profiles in order to check whether there is any dependence of the shift on wavelengths. The Mg I] 4571 \AA ~, at that phase, is noisy which makes any conclusion about a possible blueshift. 

 This behaviour was also seen in SN 1993J, and was interpreted as an indication of large scale asymmetries and clumpy distribution of the ejecta (Spyromilio. 1994)\nocite{Spy94}. The [Ca II] 7307.5 \AA~line shows a blue shift the magnitude of which seems to decrease from day 90 to day 358. Since this interval embraces the epoch of dust formation (Section 2.2) and spectral observations are so sparse it is not possible to disentangle the two possible competing effects. This scarcity of data also inhibits a more detailed discussion of the blueshift seen in the O I 7774 \AA~line at day 90. The strength of the O I 7774 \AA~feature, on the other hand, arising as it does from high levels in
an envelope that should be relatively cool, could indicate non-thermal
electron excitation analogous to what has been proposed for the He I lines at
an earlier phase.

The ``early-time'' blueshifts affecting several lines have been reported for SN 1996N by Sollerman et al. (1998)\nocite{Soll98} and extensively discussed there. Similar effects have also been reported for SNe 1987A, 1999em and 1988A. Various hypothesis have been reviewed without firm conclusions being drawn although large scale asymmetries are favoured. In spite of their reported lack of temporal evolution of the blueshifts in SN 1996N, evolution was evident in SN 1987A and seems evident in our sparse observations of [O I] 6300,64 \AA~prior to dust formation. We consider a decrease in line optical depth, possibly combined with different excitation conditions for different lines, as the most promising explanation.

On the other hand, we note the presence of fine structures superimposed clearly on the profiles of O I 7774 \AA~ (90 d spectrum), [Ca II] 7307.5 \AA~ (90 d and 298 d) and  Ca II 8662 \AA. In Figure 5 two structures on the top of [Ca II] 7307.5 \AA~and Ca II 8662 \AA~ are marked. The bumps have comparable velocities (-800 and 500 km s$^{-1}$) in both lines. Similar bumps have been reported in the 1993J spectrum around day 154 (since maximum) on the top of [O I] 6300,64 \AA ~and O I 5577 \AA ~lines, and have been attributed to asymmetry, whereas the lack of similarity in line profiles of permitted and forbidden transitions of the same element is a result of spatially different excitation conditions (Spyromilio. 1994). In the longer wavelength lines fringing in the CCD could complicate and confuse a clearer interpretation. 

The situation of SN 1990I might be more complicated in view of further information from the photometric evolution (Section 2.2), where the coexistence of both large scale asymmetries and dust condensation at later epochs (after day 237) is more probable. In fact a second blueing effect was reported in SN 1987A spectra, after $\sim$ 500 d, where the emission line profiles of Mg I] 4571 \AA~, [O I] 6300,64 \AA~ and [C I] 9824,9850 \AA~ became asymmetric with peak emission blueshifted by about $500-600$ km s$^{-1}$, and was attributed to extinction by dust within the metal-rich ejecta (Danziger et al. IAUC 4746\nocite{Dan47}, Lucy et al. 1991\nocite{Lucy}). An analysis of SN 1999em data has indicated a similar conclusion about dust formation around day 500 (Elmhamdi et al. 2003)\nocite{Elmha03}.
 
The [O I] 6300 \AA ~profile in the last two spectra displays a depletion of emission in the $[-2500:0]$ km s$^{-1}$ region. In Figure 6 we present, at three different phases of evolution, the [O I] 6300 \AA ~profile of SN 1990I compared to SNe 1993J and 1998bw. We shift arbitrarily the peaks to match each other for the clarity of the comparison. There are some points of interest on this figure. First, SN 1990I shows a broader profile  with respect to SN 1993J, and is similar to that of SN 1998bw. Second, the depletion of the blue wing is easily visible in the bottom box, pointing to different excitation conditions compared to that in the other two objects.   
Again a clumping of the radioactive material in the oxygen shell with velocities belonging to $[-2500:0]$ km s$^{-1}$ zone is a possible explanation.

Finally, the presence of the depletion structure in the 298 d and 358 d [O I] profiles prevents us from checking whether the blueshift seen in the 258 d [O I] emission increases or not with time.
\begin{figure}
\includegraphics[height=8.5cm,width=8cm]{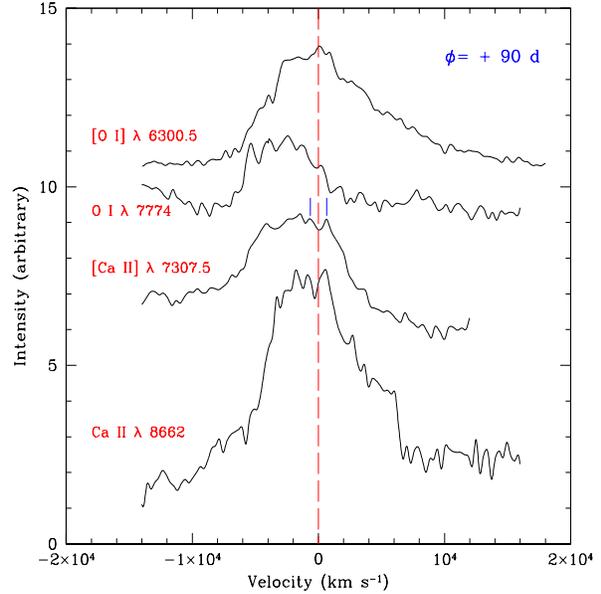}
\caption{The strongest lines, in velocity space, of the 90 d spectrum. The features are clearly asymmetric and flat topped, with the presence of fine structures. Two bumps with similar velocities in the [Ca II] 7307.5 \AA~ and Ca II IR-triplet lines are marked.}
\end{figure}
\subsubsection{Oxygen mass}
Estimating the oxygen mass is important as it can be indicative of the core mass and hence related to the nature of the progenitor star. The [O I] 6300,64 \AA~doublet line emission is known to be the primary coolant in the ejecta of Ib/c events, at least at late epochs, in view of the absence of hydrogen in this class of objects as opposed to type II events (Uomoto 1986\nocite{Uom86}; Fransson 1987)\nocite{Fran87}.

In principle, the mass of ejected oxygen can be recovered using the [O I] 6300,64 \AA~flux. Uomoto (1986) has shown that the oxygen mass, in M$_\odot$, is given by:

\begin{equation}
M_{Ox} = 10^{8} \times D^2 \times F(\rm{[O ~I]}) \times \exp{(2.28/\it T_4)}
\end{equation} 
where $D$ is the distance to the supernova (in Mpc), $F$ is the [O I] integrated flux (in ergs s$^{-1}$ cm$^{-2}$) and $T_4$ is the temperature of the oxygen-emitting gas (in 10$^4$ K). Equation 1 holds for the high density limit ($N_e \geq 10^6 ~$cm$^{-3}$) where the density is above the critical density for the [O I] line ($\sim7~\times 10^5 ~$cm$^{-3}$). This condition, on the other hand, is clearly achieved in the ejecta of type Ib/c SNe. In fact a rough estimate of the mass density can be computed at day 237. 
We have attempted to fit the [O I] 6300,64 \AA~feature with 2 Gaussians, fixing
the wavelength separation of the doublet and with both having the same velocity width. It is not possible to obtain a particularly good overall fit since one is left with either a residual in the blue wing of [O I] 6300 \AA~for narrower lines, or little evidence of the [O I] 6364 \AA~ feature for broader lines. Thus a one-component Gaussian fit may not represent the real situation where
clumping is present. On the other hand by assuming that there is a
contribution from an Fe II 6239 \AA~emission, and allowing its velocity width to be a free parameter we obtain the fit shown in Fig. 7. The $FWHM$ for the [O I] lines is $\sim 60$ \AA~ while for the Fe II 6239 \AA~line we obtain $\sim 47$ \AA. We also show
residuals which might represent the effects of clumping in the [O I] emitting
region. Conveniently, in this fit the velocity width for the Fe II
feature is less than that for [O I] as might be expected. We adopt the
velocity widths for [O I] determined in this way but note that similar values
would be obtained if we accepted the best fit of the [O I] doublet alone
around the peak emission of both components of the doublet.
\begin{figure}
\includegraphics[height=10.5cm,width=9cm]{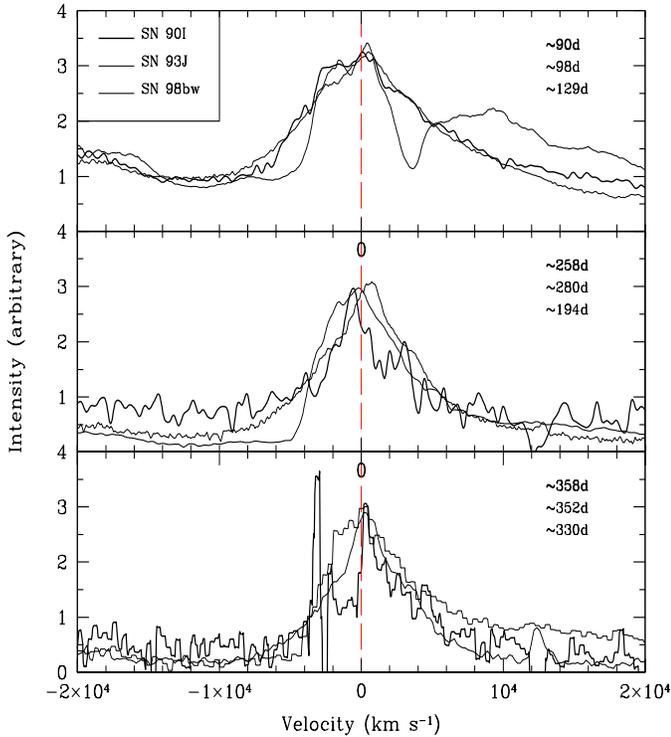}
\caption{The [O I]6300,64 \AA~line profile of SN 1990I in three different epochs of evolution (since maximum light). Comparison with SNe 1993J and SN 1998bw is made, indicating the high velocity nature of SN 1990I and the presence of a ``$depletion$'' structure in the blue edge of the emission profile (lower panel).}
\end{figure}

\begin{figure}
\includegraphics[height=9.5cm,width=9cm]{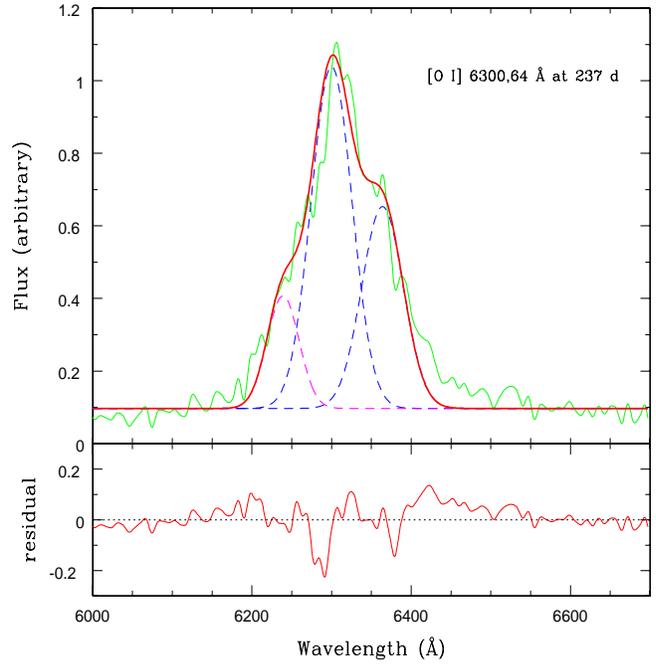}
\caption{Fit to the 237 d [O I] 6300,64 \AA~line profile of SN 1990I. Shown in thick line is the sum of the three Gaussians for [O I] 6300 \AA~, [O I] 6364 \AA~and Fe II 6239 \AA~(dashed lines; see text). The residuals are displayed in the lower panel.   }
\end{figure}

Using then the derived $FWHM$ of the [O I] line at 237 d ($\sim$ 2860 km s$^{-1}$), and adopting the ejecta mass of $M_{ej}=3.7~$M$_{\odot}$ (Sect. 3.2) we derive a value of $N_e\sim 1.16\times 10^{7}$ cm$^{-3}$. We have assumed abundance fractions of He, O and heavier elements in the following proportions: 0.1, 0.22 and 0.68, respectively. We also assume that half of the He and O and all of the other elements are singly ionized. Since we assume matter uniformly distributed this should be a lower limit for regions where clumping occurs.
 
A more accurate estimate of the density is based on the relative strengths of the [O I] doublet components (Leibundgut et al. 1991\nocite{Lei91}; Spyromilio $\&$ Pinto. 1991\nocite{Spy91}). The authors of both papers have computed the variation of the line doublet ratio as function of the density at a given time and found it insensitive to the adopted temperature, especially at late epochs. From the 237 d [O I] line profile fit we measure a relative intensity $I(6300)/I(6364)\simeq1.69$, similar to that of SN 1987A at similar phase ($\sim$ 1.45 around day 250). The uncertainty in the temperature leads to the following range $2.3\times 10^{9} \leq N_{e} \leq 4\times 10^{9}$ according to Fig. 6 of Leibundgut et al. (1991). The large difference (a factor of about 100) between this density estimate and the one from the simple approach using the volume calculated from the [O I] velocity and M$_{ej}$ indicates, indeed, that the oxygen material fills the volume at day 237 in a clumpy way rather than homogeneously. This high density would favour the presence of Fe II features mentioned previously and identified in Figure 1. 
 
Equation 1 implies time-dependence because of the variation of $F$([O I]) and the temperature. Schlegel $\&$ Kirshner (1989)\nocite{Schl89}, when estimating the oxygen masses ejected by SNe Ib 1984L and 1985F using this method, have adopted a constant temperature $T_4$=0.4 at the nebular phase. The assumption of a constant temperature at different late phases provides different oxygen masses as one may expect, since earlier nebular epochs are hotter compared to latter ones. Worth noting is the exponential dependence of the oxygen mass on temperature.

A method for estimating temperature at a given time is based on the [O I] 5577 \AA~ to [O I] 6300,64 \AA~flux ratio. This comes from the assumption that the O I lines are formed mainly by collisional excitation, and the ratio is given by the following expression (Houck \& Fransson 1996)\nocite{Hou96}:   

\begin{equation}
\frac{F(6300)}{F(5577)}=0.03 \beta_{6300} \times [1+1.44 \it T_3^{-0.034}(\frac{10^8}{N_e})] 
\end{equation}
\vspace{-0.3truecm}\hspace{2truecm}$\times \exp{(25.83/ \it T_3)} $

\vspace{0.3truecm}\hspace{-0.6truecm}where $T_3$ is the temperature of the oxygen-emitting gas (in 10$^3$ K), $\beta_{6300}$ is the [O I] 6300 \AA~Sobolev escape probability ($\simeq$ 1) and $N_e$ is the electron density. 

Because of line blending, mainly with Fe II lines, the [O I] 5577 \AA~flux measurements are less precise, especially for the last three spectra. We use the 237 d spectrum because it is less noisy and does not show the depletion feature seen later in the blue side of the [O I] 6300,64 \AA~emission profile. 

A lower limit on the ratio (left hand term in Eq. 2) can be estimated using an upper limit on the [O I] 5577 \AA~flux by integrating over the same velocity interval of the [O I] 6300,64 \AA~profile. 

When using the density range estimated on the basis of the [O I] 6300,64 \AA~doublet ratio we derive temperatures of the order of $3200-3500$ K. We obtain then $0.7-1.35$ M$_\odot$ as oxygen mass. Here we note that since we are adopting an upper limit on the [O I] 5577 \AA~flux, the recovered oxygen mass is a lower limit. 
The very weak [O I] 5577 \AA~ feature in the last three spectra is an indication of low temperatures in the oxygen-emitting zone. Indeed, although the determination of oxygen mass following this method suffers from several uncertainties, it seems that SN 1990I has reached lower temperatures earlier and has a higher oxygen mass compared to typical Ib/c SNe. In Table 2 we report the amount of ejected oxygen mass for a selected sample of type Ib/c SNe.   
\begin{table}
\begin{minipage}{80mm}
 \caption{The estimated ejected oxygen mass.} 
\bigskip
\centering
\begin{tabular}{ccccc}
\hline \hline
SN  & SN&$M_{Ox}$ &References  \\
name &type &$($M$_{\odot})$ & \\
\hline
1987A& II&1.5-2& 1, 2\\
1984L& Ib&$\sim 0.3$$^\star$ & 3 \\
1985F& Ib&$\sim 0.3$$^\star$& 3 \\
1987M& Ic&0.4$^\star$ & 4 \\
1990I& Ib&0.7-1.35$^\star$  &this work \\
1993J& IIb & $\sim 0.5$&5  \\
1996N& Ib & 0.1-0.3$^\star$ &6 \\
\hline \hline
\end{tabular}
\end{minipage}\\
\\
{\emph{{\rm \scriptsize
Refs:\\1- Fransson et al. 1993; 2- Chugai 1994; 3- Schlegel $\&$ Kirshner 1989; \\4- Filippenko et al. 1990; 5- Houck $\&$ Fransson 1996; 6- Sollerman et al. 1998.  
\\$\star$ Estimated adopting the same method (i.e. using Eq. 1)}}}\\ 
\normalsize
\end{table}
The table includes as well data of SN II 1987A and SN IIb 1993J. One should note however that adopting a different method, Tutukov $\&$ Chugai (1992)\nocite{Tut92} have argued for an amount of $\sim 0.8~$M$_{\odot}$ as total ejected oxygen mass\footnote{$0.2~$M$_{\odot}$ neutral and $0.6~$M$_{\odot}$ as ionized-oxygen mass.} (neutral and ionized) for type Ib SN 1984L, and concluded that a typical value of $\sim 1~$M$_{\odot}$ should be ejected as well by SNe Ib 1983N and 1985F. 

The somewhat higher oxygen mass and lower temperature with respect to other SNe Ib/c may be related to other  observational peculiarities in SN 1990I. First the higher expansion velocity (i.e. faster cooling), and second the possible dust condensation at a time as early as $\sim$250 d (be deduced photometrically later).   
\subsection{Photometric evolution}
The photometry data of SN 1990I are reported in Table 3 together with the different instruments with which the observations were obtained. The SN magnitudes have been measured with the ROMAFOT package in MIDAS (Buonanno et al. 1983)\nocite{Buon83}. After measuring the point spread function from the field stars, the programme allows the simultaneous fitting of the SN and the background, which has been approximated by a tilted plane. 
In Figure 8, the B, V, R and I light curves are displayed. The well sampled light curves, following maximum, coupled with the reported early spectral properties (i.e. broad P-Cygni profiles superimposed on a blue continuum) provide the possibility of estimating the time of maximum light and roughly the explosion date. In fact  
based on the shape of the light curves during the first $\sim$ 30 days of evolution and bearing in mind that the typical rise time of type Ib/c SNe is about $15-20$ days to reach maximum light, we estimate both the date of maximum light and the explosion time to be, respectively, April 29 (1990; JD 48010$\pm$ 5d ) and April 11 (1990; JD 47992 $\pm$ 5d). These dates are taken as references through this work. However, because the point of maximum was not defined (see Fig. 8) we estimate that maximum light could have occured at most 5 days earlier.

\begin{table*}
\centering
\begin{minipage}{140mm}
  \caption{ SN 1990I Photometric Observations}
\begin{tabular}{c c c c c c c c}
\hline \hline
 Date &      JD   &  B($\sigma$) & V($\sigma$) & R($\sigma$) & I($\sigma$)& Instrument\\
(UT) &2400000+&&&&&&\\
\hline        
27/04/90 &    48008.5 &    --- &   15.42 .03 &     --- &  --- &	ESO-2.2m \\
29/04/90  &    48010.5  &  16.68 .05   &  15.58 .03  &  ---   &---&ESO-2.2m \\
30/04/90  &    48011.5  &  16.75 .05   &  15.65 .03  &   ---  & ---&ESO-2.2m \\
01/05/90  &    48012.5  &  ---   &  15.73 .03  &   ---  & ---&	ESO-2.2m \\
02/05/90  &    48013.5  & 17.01 .05    &  15.80 .03  &  15.28 .03  & ---& CTIO-0.91m\\
10/05/90  &    48021.5  &   17.35 .05  &  16.18 .03  &  15.60 .03  & ---&ESO-NTT\\
14/05/90  &    48025.5  &  17.65 .05   &  16.44 .03  &  15.87 .03  & 15.38 .03&CTIO-0.91m\\
18/05/90  &    48029.5  &  ---   &  16.50 .03  &  ---  & ---&ESO-NTT\\
28/05/90  &    48029.5  &  17.85 .05   &  16.78 .03  & ---   &---&ESO-NTT\\
08/06/90  &    48050.5  & 18.18 .05    &  17.06 .03  &  16.55 .03  & 15.96 .03&CTIO-0.91m\\
12/06/90   &   48054.5  & 18.22 .05    &  17.16 .03  &  16.64 .03  & 16.05 .03 &CTIO-0.91m\\
16/06/90   &   48058.5  & 18.26 .05    &  17.23 .03  &  16.71 .03  & 16.12 .03&CTIO-0.91m\\
19/06/90   &   48061.5  & 18.27 .05    &  17.26 .03  &  16.75 .03  & 16.13 .03&ESO-1.54m\\
20/06/90   &   48062.5  &   18.24 .05  &  17.27 .03 &16.72 .03  & ---&ESO-1.54m\\
21/06/90   &   48063.5  &  18.34 .05   &  17.30 .03  &  16.83 .03  & 16.25 .03&CTIO-4.0m\\
25/06/90   &   48067.5  &  18.40 .05   &  17.37 .03  &16.92 .03  & 16.32 .03&CTIO-4.0m\\
29/06/90   &   48071.5  & 18.44 .05    &  17.43 .03  &  16.97 .03  & 16.37 .03&CTIO-4.0m\\
04/07/90   &   48076.5  &  18.53 .05   &  17.58 .03  & 17.09 .03 &  16.44 .03&CTIO-0.91m\\
12/07/90   &   48084.5  & 18.63 .05    &  17.71 .03  &   17.20 .03 &  16.56 .03&CTIO-0.91m\\
26/07/90   &   48098.5  &   18.78 .05  &  17.94 .03  & 17.42 .03  & ---&ESO-3.6m\\
01/08/90   &   48104.5  & 18.98 .05    &  18.15 .03  &  17.55 .03  & 16.92 .03&CTIO-0.91m\\
12/08/90   &   48115.5  & 19.11 .05    &  18.30 .03  & 17.73 .03  & 17.08 .03&CTIO-0.91m\\
22/08/90   &   48125.5  & 19.32 .05    &  18.49 .03 & 17.86 .03  & 17.26 .03&CTIO-0.91m\\
23/11/90  &   48218.5  &  ---   &  20.47 .1  & --- &  ---&CTIO-0.91m\\
11/12/90  &   48236.5  &  22.26 .3   &  21.46 .1  & 19.98 .03 &  19.82 .03&CTIO-4.0m\\
14/12/90  &   48239.5  &   ---  &  21.11 .1  &20.04 .1  & ---&ESO-1.54m\\
15/12/90  &   48240.5  &   ---  &  21.32 .1  &  --- &  19.72 .03&CTIO-0.91m\\
21/12/90  &   48246.5  & ---    &  21.34 .1  & ---  & ---&ESO-3.6m\\
28/12/90  &   48253.5  & 22.42 .3    &  21.88 .1  &  20.60 .1 &  20.11 .1&CTIO-0.91m\\
15/01/91  &   48272.5  &  ---   &  ---  &   21.02 .1 &  ---&CTIO-0.91m\\
16/01/91  &   48273.5  & ---    &  22.42 .2  &   21.15 .1 &  20.99 .1&CTIO-0.91m\\
12/02/91  &   48300.5  & 23.28 .35   &   23.01 .25  &   21.70 .1 &  21.40 .1&CTIO-4.0m\\
20/02/91  &   48308.5  &  22.90 .35  &   22.90 .25  &  21.85 .1  & 21.56 .1& ESO-3.6m\\
09/04/91  &   48356.5  &  ---  &   ---  & 23.37 .25  & --- &CTIO-4.0m\\

  \hline \hline\\
\end{tabular} 
\end{minipage}
\normalsize
\end{table*}

SN 1990I suffers high galactic extinction, $A_V^{gal}=0.374$ mag, according to maps of the galactic dust distribution by Schlegel, Finkbeiner $\&$ Davis.(1998). This corresponds to a colour excess about $E(B-V)=0.12$, where the standard reddening laws of Cardelli et al. (1989)\nocite{Card89} have been used. In addition we note the presence of a narrow NaI D interstellar absorption feature in the 12 d spectrum at the redshift of the parent galaxy, with an equivalent width $(EW)\sim$0.26 \AA. This can be used to estimate the reddening in the host galaxy. Using correlations relating EW(Na I D) to the reddening within the parent galaxy from the literature we obtain $E(B-V)\sim0.06$ (Barbon et al. 1990)\nocite{Barb90} and $E(B-V)\sim0.034$ (Benetti et al., in preparation). In what follows, we adopt $A_V =0.4$ mag as an estimate of the total extinction.

After the fast fall from maximum light ($\gamma_V\sim$5.6 mag (100 d)$^{-1}$), SN 1990I settles onto a linear decline with a rate in the V band $\gamma_V\sim$1.92 mag (100 d)$^{-1}$ up to about 230 days since explosion (dotted line in Fig. 8). The $^{56}$Co to $^{56}$Fe decay slope, 0.98  mag (100 d)$^{-1}$, corresponding to full $\gamma$-ray trapping is also shown. Except for rare cases where the late slope approaches the full trapping rate (e.g. SN Ib 1984L; Schlegel $\&$ Kirshner 1989), the steeper decline rate of about 1.9 mag (100 d)$^{-1}$ is common for type Ib/c SNe at late epochs (e.g. SN Ib 1983N, SN Ic 1983V, SN Ic 1994I and SN IIb 1993J, Clocchiatti $\&$ Wheeler 1997\nocite{Cloc97}; SN Ib 1996N, Sollerman et al. 1998). This rapid decline, on the other hand, is indicative of significant $\gamma$-ray escape resulting from a low mass ejecta in this class of object. 
\begin{figure*}
\includegraphics[height=13.5cm,width=16cm]{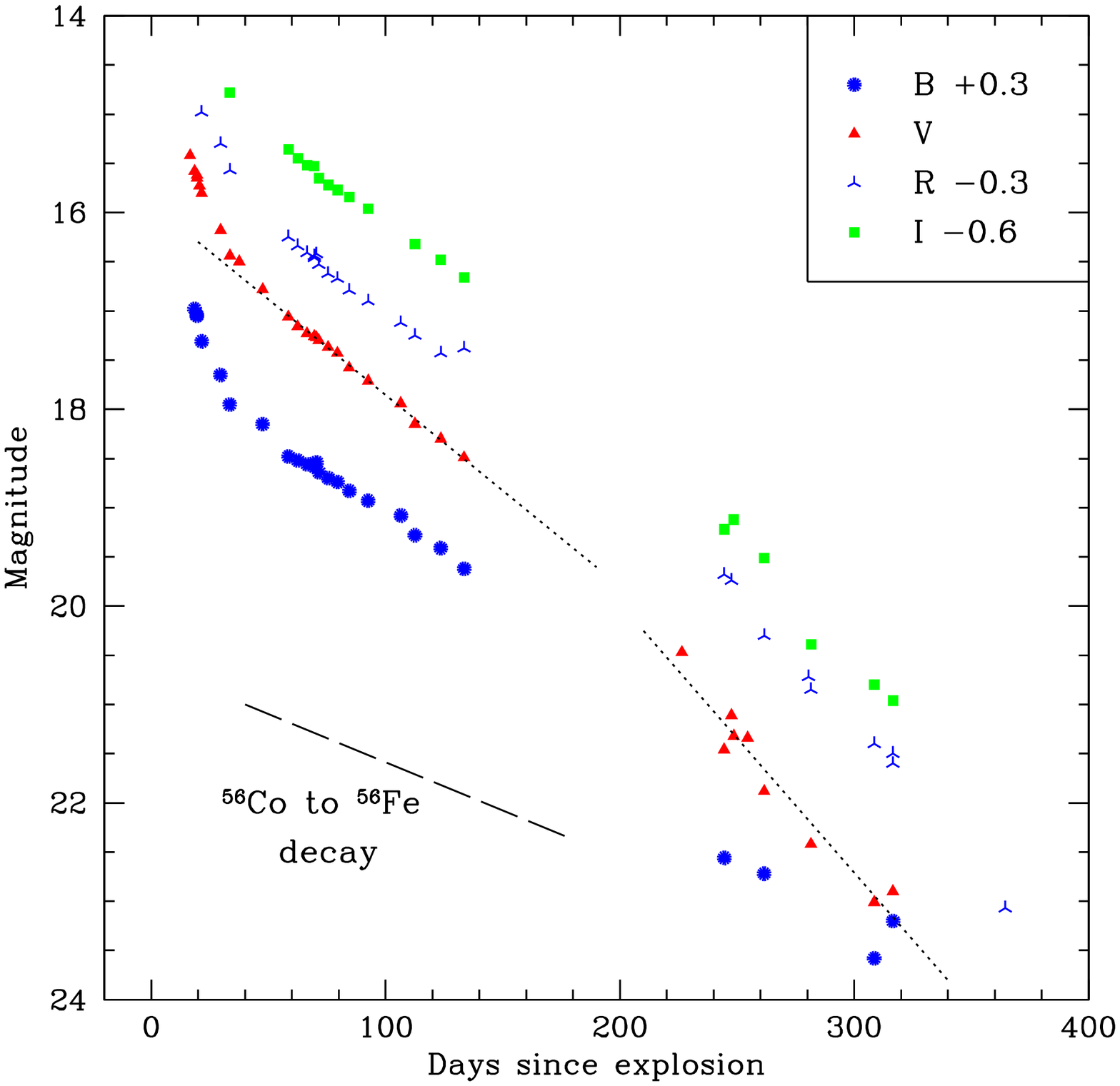}
\caption{ B, V, R, and I light curves of SN 1990I. The light curves have been shifted by the reported amounts. Also shown are the fits to the V-light curve tail (dotted lines) and as well the $^{56}$Co to $^{56}$Fe decay slope (long-dashed line).}
\end{figure*}

Interestingly, however, SN 1990I shows an even steeper decline starting at about day 230 with a V-band decline rate $\gamma_V\sim$2.7 mag (100 d)$^{-1}$. In addition, while the R and I light curves confirm this behaviour, the B light curve tends to flatten. The least-squares fit to the light curves at the three different phases of evolution are reported in Table 4. At early phase, the B light curve declines faster compared to the other bands, whereas for the two following phases it exhibits a lower decline rates. The V band, on the other hand, is the faster fading band during the following two phases. We compute as well the third to second phase decline slope ratio for each band. Results are shown in Table 4. The change in slope behaviour at late phases is seen also in other events, namely SN 1993J, SN 1998bw and SN 1996N with different occurrence times. SN 1996N, between 184 d and 312 d, seems to have similar V, R and I decline rates as SN 1993J (Sollerman et al. 1998). The last photometric point of SN 1996N (around day 342) is clearly deviating from the linear trend of the previous data, indicating the occurrence of a phase with steeper decline after day 312. In Table 4 we present the SN 1996N third to second decline rate ratios using the available photometric data from Sollerman et al. (1998). Interesting points can be read out from Table 4. A comparison of ratio values demonstrates the peculiar behaviour of SN 1990I with an increasing ratio towards the red bands, indicating some kind of dependence of the defined ratio on wavelength. Although there is a paucity of SN 1996N data, it is similar to SN 1990I in having a ``$greater~ than ~one$'' ratio, whereas SNe 1993J and 1998bw tend to flatten. Furthermore, SN 1993J has an opposite behaviour with respect to SN 1990I, in the sense that its B light curve declines more rapidly in the third phase while the V, R and I light curves flatten. The opposite holds for SN 1990I.

\begin{figure*}
\includegraphics[height=13.5cm,width=16cm]{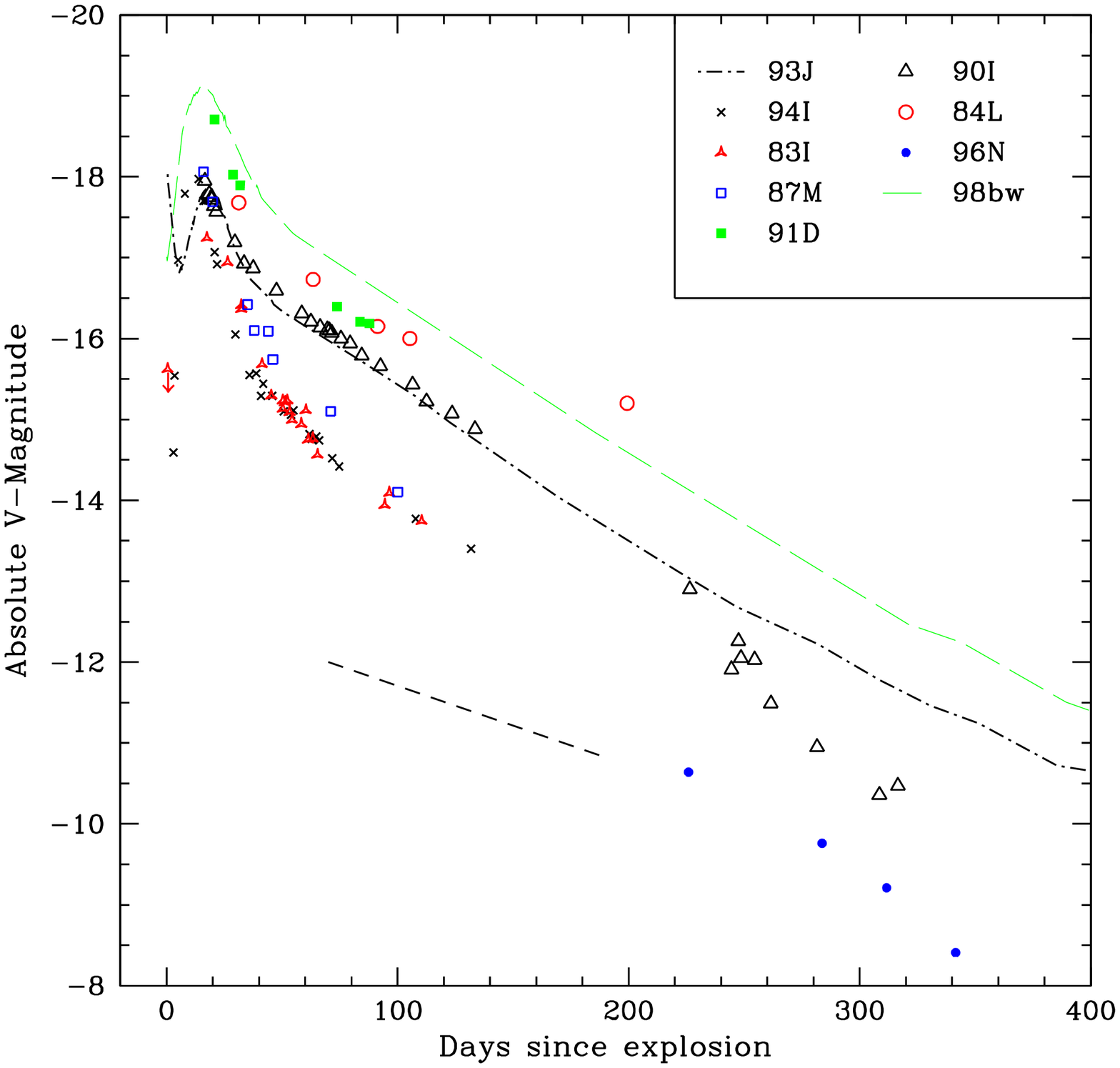}
\caption{Comparison of the V absolute light curve of SN 1990I with those of other Ib/c events. The $^{56}$Co to $^{56}$Fe decay slope is shown (short-dashed line). The used parameters are reported in Table 5. }
\end{figure*}
\begin{table*}
\begin{minipage}{100mm}
 \caption{Decline rates of SN 1990I. Data for events with two slopes at late times are also reported (see text for discussion).} 
\bigskip
\centering
\begin{tabular}{cccccc}
\hline \hline

Phase range  & $\gamma_B$ & $\gamma_V$ & $\gamma_R$ & $\gamma_I$  \\
(days) & mag (100 d)$^{-1}$ & mag (100 d)$^{-1}$ & mag (100 d)$^{-1}$ & mag (100 d)$^{-1}$  \\

\hline

$[16:40]$   & 6.2  & 5.6 & 4.78 & $---$\\

$[50:150]$  & 1.53 & 1.92 & 1.77 & 1.7 \\

$[200:340]$ & 1.2 & 2.7 & 2.6 & 2.6\\
\hline
The third to second phase decline slopes ratio~:\\
\hline
$\gamma_{[200:340]}$/$\gamma_{[50:150]}$(90I) & 0.79 & 1.4 & 1.467 & 1.53\\
$\gamma_{[150:350]}$/$\gamma_{[45:150]}$(93J) & 1.47 & 0.82 & 0.62 & 0.75\\
$\gamma_{[310:490]}$/$\gamma_{[40:330]}$(98bw) & 0.86 & 0.56 & 0.78 & 0.745\\
$\gamma_{[t \geq 312]}$/$\gamma_{[t\leq 312 ]}$(96N) & $---$ & $\sim$1.6 & $\sim$1.4 & $---$\\
\hline \hline
\end{tabular}
\end{minipage}\\
\\
\normalsize
\end{table*}
In Figure 9 the absolute V light curve of SN 1990I is compared with other type Ib/c core collapse events. For SN 1990I a heliocentric velocity corrected for the Local Group infall onto the Virgo Cluster is 2751 km s$^{-1}$ (as reported in the LEDA database) translates into $\mu \sim 32.97$. Table 5 summarizes the main parameters adopted for the selected sample.
The figure reveals a strong similarity between SN 1990I and 1993J, from maximum to about day 230, with SN 1990I about 0.2 mag brighter. The maximum brightness is, on the other hand, quit comparable to those of SNe 1987M and 1994I and intermediate between the bright SNe 1999bw and 1991D and the faint SN 1983I. Moreover SNe 1990I, 1993J, 1991D and 1998bw decline from maximum with a similar rate to reach the exponential decay and indicate similar peak-to-tail contrast, whereas SNe 1983I, 1987M and 1994I display narrower peak widths and greater peak-to-tail contrast. The sample we are studying, although small, shows a correlation between the peak width and the peak-to-tail contrast. SN 1984L seems to be the extreme case of the sample, with the smaller peak-to-tail contrast and a possibly broader peak. The decline rate from maximum and the peak-to-tail contrast are computed and shown in Table 5 for events with well sampled early data. Interestingly, the events with steeper decline rates (i.e. narrow widths) and greater peak-to-tail contrast are classified as type Ic SNe, while the remaining events are type Ib. This is not surprising since the width and the way the light curve declines from maximum during this early phase of evolution are related to the progenitor star properties, namely the mass, radius, energy and the degree of mixing. These parameters, especially the ejected mass, controls the time the trapped decay energy takes to diffuse through the envelope (Ensman $\&$ Woosley 1988)\nocite{Ens88}. In fact, for smaller mass the maximum is reached earlier because of larger radioactive heating and higher expansion velocity. The smaller the ratio ``$M_{ej}/E$'' and the closer the $^{56}$Ni to the surface the faster the light curve declines (Shigeyama et al. 1990)\nocite{Shi90}. One scenario for SNe Ic shows, indeed, that they originate from slightly smaller masses at the time of explosion compared to type Ib objects (Yamaoka $\&$ Nomoto. 1991)\nocite{Yam91}. If however SNe Ic have been stripped of helium the main sequence progenitors could have larger masses than those of type Ib SNe.

As discussed before, and except for SN 1984L, all the SNe of the sample decay, on the exponential tail, faster than $^{56}$Co (short-dashed line in Figure 9). If we assume the V light curves trace approximately the bolometric ones, they all reflect smaller ejecta masses allowing greater $\gamma-$ray escape. After $~$230 d, SN 1990I deviates clearly from its previous slope and the deviation from the SN 1993J tail increases with time, approaching even the SN 1996N light curve. In fact, while around day 225 the SN 1990I light curve coincides with SN 1993J tail a deviation about 1.3 mag is measured around day 310. 
We note the rapid change seen in the $B-V$ colour evolution at the same age. Figure 10 reports the $B-V$ intrinsic colour of~ SN 1990I together with that of other type Ib/c events for comparison. $(B-V)_0$ of SN IIb 1993J is also shown (dashed line). The comparison indicates that SN 1990I follows a similar evolutionary trend as the other SNe, displaying a rapid reddening during the first 40 days, reflecting the cooling due to the envelope expansion. After a short duration peak it turns to the blue with a slope in the $[40:120]$ days time range of about 0.43 mag (100 d)$^{-1}$, similar to SNe 1984L, 1987M and 1991D, although it is about 0.2 mag bluer. At this epoch, SN 1993J exhibits a faster decline ($\sim$ 0.92 mag (100 d)$^{-1}$). The colour peak, around day 40, corresponds to the beginning of the exponential radioactive tail in the V-absolute light curves (Fig. 9). Then in Fig. 10, the SNe settle on an almost $``plateau''$ phase of long duration, well sampled for SNe 1993J and 1990I. Unfortunately, late observations of SNe 1987M and 1991D are not available, however late phase data of SNe 1984L and 1985F may indicate the presence of a plateau phase.       

\begin{figure}
\includegraphics[height=9.5cm,width=9cm]{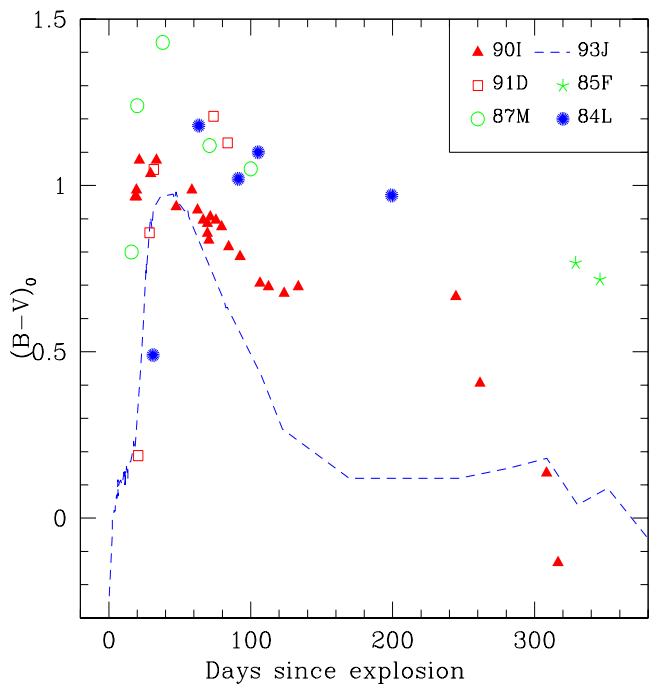}
\caption{The evolution of the $B-V$ intrinsic colour of SN 1990I, compared to those of other Ib/c events. Adopted reddening for each event is reported in Table 5. }
\end{figure}
Again, the SN 1990I intrinsic $(B-V)$ colour displays a dramatic fall after day 245. It decreases by $\sim$0.8 mag in about 75 days. The time of this rapid and steep decline reflects the occurrence of the previously seen deviation in the V-absolute light curve. These facts point to the presence of a phenomenon able to cause a deficit in the optical light curves and also to cause the line profiles to be blueshifted in a short time ($\Delta t/t \sim$ 0.1; Section 2.1). This rapid transformation of profiles during the nebular epoch has also been reported in SNe 1987A and 1999em, and when combined with a sudden increase in the decay rate of the optical light curves, it has been interpreted as an indication of dust formation. We believe that this the case for SN 1990I also. In fact, bearing in mind that dust condensation is expected to correlate with the cooling of the ejecta, we interpret the formation of dust earlier in SN 1990I compared to SNe 1987A ($\sim$530 d) and SN 1999em ($\sim$500 d) to be due to lower ejecta mass and rapid cooling for this event. It might also result from the higher metallicity content in the envelope of SN 1990I. SN 1987A shows a slight wiggle around day 550 in the optical colour evolution ($B-V$) (Suntzeff $\&$ Bouchet 1991\nocite{Sun91}). That was interpreted as a blueing associated with dust formation, but was not so dramatic as demonstrated by SN 1990I. Such a blueing effect at the time of dust formation has been discussed quantitatively in the case of SN 1987A by Lucy et al. (1991). The authors suggest that the blueing, reported in both ($B-V$) and ($U-B$) colours at the onset of dust condensation, resulted from an increase of dust albedo at shorter wavelengths. In fact one knows from the work of Draine (1985) that for both graphite and astronomical silicates the albedo does increase with decreaing wavelength in the optical range. Another explanation, in the case of SN 1987A, has been provided by Phillips $\&$ Williams (1991)\nocite{PhilW}, ascribing the increasing flux in the blue to the increasing contribution from [Fe II] lines, however even including all available Fe lines, Danziger et al. (1991) did not succeed in getting such a strong effect. The blueing at the time of dust formation could be a result of both dust albedo effect and emission lines (both Fe and others). In any case the ``dramatic'' blueing seen in SN 1990I depends quantitatively on the accuracy or reliability of the ($B-V$) measurement at day 240. Modelling are required to quantitatively check this effet.

\begin{table*}
\begin{minipage}{100mm}
 \caption{Main parameters of the SNe sample.} 
\bigskip
\centering
\begin{tabular}{ccccccccc}
\hline \hline
SN  & SN & Host & Distance& $A_V$ &$\gamma_V$$^\ddag$ & Max-to-tail contrast &References  \\
name & type & galaxy &(Mpc)&(mag)&(mag (100 d)$^{-1}$)&(mag) \\
\hline
1990I & Ib & NGC 4650 & 39.3$^\star$ & 0.4 &$\sim$5.7 &$\sim$1.05& this work \\
1983I & Ic & NGC 4051 & 13.17$^\star$ & 0.043 &$\sim$8.4&$\sim$2.2 & 1, 2\\
1984L & Ib & NGC 0991 & 20.44$^\star$ & 0.0 &$---$&$---$&3, 4, 5 \\
1987M & Ic & NGC 2715 & 22.7$^\star$ & 1.3 & $---$&$---$&6\\
1991D & Ib & PGC 84044 & 178.52$^\star$ & 0.186 &$---$&$---$& 7\\
1993J & IIb & M81 & 3.64 & 0.93 &$\sim$7&$\sim$1.23& 8\\
1994I & Ic & NGC 5194 & 8.32 &1.4 &$\sim$11 & $\sim$2.42& 9\\
1996N & Ib & NGC 1398 & 22 & 0.0 &$---$&$---$& 10\\
1998bw & Ic-Hypernova & ESO 184-G82 &35.16$^\star$ & 0.2&$\sim$6&$\sim$1.64& 11, 12, 13, 14\\
\hline \hline
\end{tabular}
\end{minipage}\\
\\
{\emph{{\rm \scriptsize Refs:\\
1- Tsvetkov 1985\nocite{Tsv85}; 2- Tsvetkov 1988\nocite{Tsv88}; 3- Meikle et al. 1984\nocite{Mei84}; 4- Buta 1984\nocite{But84}; 5- Tsvetkov 1987\nocite{Tsv87}; \\6-  Filippenko et al. 1990; 7- Benetti et al. 2002; 8- Lewis et al. 1994\nocite{Lew94}; 9- Richmond et al. 1996;\\ 10- Sollerman et al. 1998; 11- Galama et al. 1998\nocite{Gal98}; 12- McKenzie $\&$ Schaefer 1999\nocite{Mc99}; 13- Patat et\\ al. 2001; 14- Sollerman et al. 2000\nocite{Soll00}. 
\\$\star$ Estimated using ``LEDA'' database and adopting $H_0=70~ km ~s^{-1} Mpc^{-1}$.\\ 
$\ddag$ Early decline rate from maximum.}}}\\
\normalsize
\end{table*}
\section{Physical parameters estimate}
In this section we construct the bolometric light curve of SN 1990I, and adopt a very simple model for radioactive powering. A reasonable fit to the data, especially during the initial part of the radioactive tail after the steep decline from the maximum brightness, provides quantitative insights on the basic parameters responsible for the bolometric light curve shape, namely the ejecta and $^{56}$Ni masses.
\subsection{Bolometric light curve}
In Figure 11 we present the ``bolometric'' light curve of SN 1990I together with that of SN 1993J (Richmond et al. 1994, 1996)\nocite{Rich94} for comparison. The luminosities were computed by integrating the fluxes using the available photometric data bands. The conversion of the magnitudes to fluxes was carried out assuming the calibrations of Bessell (1979)\nocite{Bess79}. The adopted parameters, i.e. distance and reddening, are reported in Table 5. The derived luminosities, however, should be corrected for the near-IR ($JHK$) fluxes. To account for this contribution, in the case of SN 1998bw, a fraction of 42$\%$ at day 370 was estimated to be added to the $L_{BVRI}$ (Sollerman et al. 2002)\nocite{Soll02}, whereas Patat et al. (2001)\nocite{Pat01} have derived a fraction of about 35$\%$ on day 65.4. A fraction of about 50$\%$, on the other hand, has been estimated around day 440 for SN IIP 1999em (Elmhamdi et al. 2003). Though not precise, in view of the possible increasing importance of the near-IR fraction with time (e.g. SN 1998bw; Sollerman et al. 2002), here we simply adopt 35$\%$ as a constant percentage contribution of the near-IR fluxes. The bolometric light curves of SN 1990I and 1993J shown in Figure 11 have been indeed scaled up to account for this constant contribution.

The constructed bolometric light curve of SN 1990I can be divided into three phases. $First ~phase$ (the initial 40 days): although the rising part is not covered, the slow drop from the maximum compared to SN 1993J is confirmed. We note the similarity starting around day 30 with SN 1993J, not only in the decline rate but as well in the computed luminosity. Around day 20 SN 1990I was $\sim$0.15 dex fainter. $Second ~phase$ (the $[50:200]$ days time interval): the light curve settles on the exponential decay tail, with a match with SN 1993J at least until day 100, whereas around day 130 SN 1990I seems to be $\sim$ 0.06 dex brighter. At this epoch, an $e$-folding time of 60 $\pm$2 days is measured, considerably faster than the one of $^{56}$Co decay (i.e. 111.3 d), suggesting the $\gamma$-rays escape with decreased deposition, owing to the low mass nature of the ejecta. $Third ~phase$ (after day 200): at this phase both SNe (1990I $\&$ 1993J) tend to behave differently. While SN 1993J flatten ($e$-folding time $\simeq$85 $\pm$2 d), SN 1990I shows the dramatic fall seen in the $VRI$ broadband photometry ($e$-folding time $\simeq$47 $\pm$2.8 d). 

The observed late time flattening in SN 1993J (and also in SN 1998bw) can be explained as due to the interaction of the ejecta with the circumstellar material (Sollerman et al. 2002), although for SN 1998bw Chugai (2000)\nocite{Chug00} has found good agreement with observations adopting complete $^{56}$Ni mixing and a higher matter density in the central part of the envelope. On the other hand, the sudden drop in luminosity after day 200 combined with spectral features in the case of SN 1990I point to dust condensation. 
\subsection{Simple $\gamma$-ray deposition model}
\begin{figure*}
\centering
\includegraphics[height=10.5cm,width=14cm]{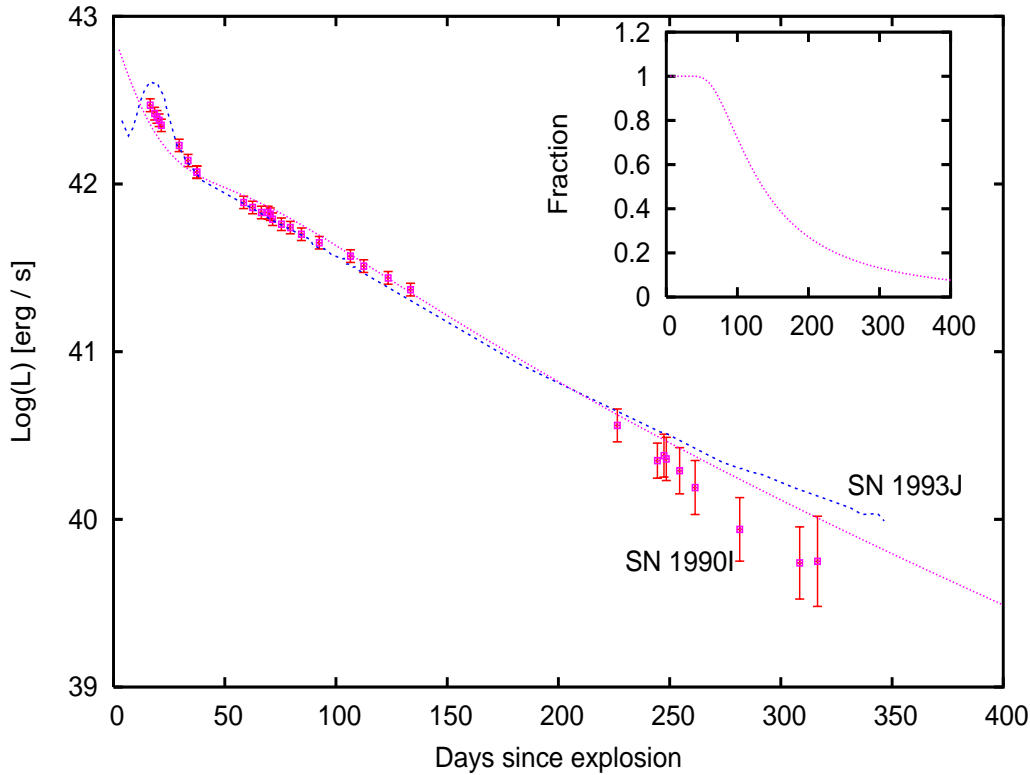}
\caption{ The computed ``$BVRI$'' bolometric light curve of SN 1990I together with that of SN 1993J for comparison. SNe 1990I and 1993J luminosities are shifted upward to account for a ``$JHK$'' contribution of the order of 35$\%$. The best fit with a simplified $\gamma$-ray deposition model from $^{56}$Ni$\rightarrow$$^{56}$Co$\rightarrow$$^{56}$Fe is shown. It corresponds to M$_{ej} \sim$ 3.7 M$_{\odot}$ and M(Ni)$\sim$ 0.11 M$_{\odot}$. The window displays the evolution of the fraction of the $\gamma$-ray luminosity deposited in the envelope corresponding to the best fit.}
\end{figure*}
The principal goal of the simple light curve modeling is to fit the bolometric behaviour starting at the transition phase (i.e. the early part of the radioactive decay tail) up to about 100 d. In the time range of interest, the contributions from radioactive elements other than $^{56}$Ni and $^{56}$Co are not taken into account, for the simple reason that the more long-lived elements, like $^{57}$Co, shown to be important at late phases (Woosley et al. 1989)\nocite{Woos89} are not sufficiently abundant to affect the light curves at day 100 (Danziger et al. 1991)\nocite{Dan91}. 

The model describes the luminosity evolution in an homologously expanding spherical ejecta with a point source $\gamma$-ray deposition from $^{56}$Ni$\rightarrow$$^{56}$Co$\rightarrow$$^{56}$Fe. It is based on the simple approach adopted by Swartz $\&$ Wheeler (1991)\nocite{Swar91} and Clocchiatti $\&$ Wheeler (1997) when studying, respectively, the late time light curve of SN 1984L and type Ib/c SNe representative light curves. In fact, assuming free expansion of the ejecta ($v(r,t) = r/t$), a power law density distribution ($\rho (r,t) \propto r^{-n}(t)$) and a purely absorptive opacity ($\kappa_\gamma$) being constant throughout the ejecta, the equations for the mass, kinetic energy and $\gamma$-ray optical depth can be solved, giving rise to an expression for the total $\gamma$-ray optical depth relating the main physical parameters (Clocchiatti $\&$ Wheeler 1997): 
\begin{equation}
\tau _\gamma = C \kappa_\gamma \times M_{ej}^2/E \times t^{-2}
\end{equation}
where t refers to time since explosion and C is a constant depending on the density function. For a power index $n=7$ we have $C\simeq 0.053$, adopted in our calculations. A typical value for the opacity, $\kappa_\gamma = 0.033 $cm$^2$ g$^{-1}$, is also assumed (Woosley et al. 1989). 

Within this simplified picture, the emergent $\gamma$-ray luminosity, at a given time t, reads:
\begin{equation}
L(t) = L_0(t) \times [1 - exp(-\tau _\gamma(t))]
\end{equation}
where  L$_0(t)$ is the total $\gamma$-ray luminosity from radioactive decay.

In order to estimate L$_0$ at a given time, a compilation of the radioactive decay energy based on the properties of the $^{56}$Ni$\rightarrow$$^{56}$Co$\rightarrow$$^{56}$Fe decay is needed. For this purpose, we adopt the main parameters presented by Jeffery (1999)\nocite{Jeff99} and Nadyozhin (1994)\nocite{Nady94}. In this scheme the total released energy via $\gamma$-photons per $^{56}$Ni decay is calculated to be 1.73 Mev, while the mean photon plus positron kinetic energy per $^{56}$Co decay is computed to be 3.74 Mev. The positron fraction in energy is only about 3.5$\%$ of the total energy per $^{56}$Co decay, while the neutrinos are assumed to escape the ejecta without any energy contribution (Jeffery 1999).

Finally, we compute the total rate of radioactive energy production at a given time and combine it with equations 3 and 4, thus providing a general description of the simple radioactive decay energy deposition model in a spherical geometry. The model, although simple, might be a good indicator for $^{56}$Ni and ejecta mass estimates, especially at the transition phase (i.e. in the $[40:100]$ days time range).   

Figure 11 displays the best fit of SN 1990I bolometric light curve (dotted line). Adopting, in fact, an energy $E(10^{51} $ergs$)=1$, our best model fit indicates the following values: $M(^{56}Ni)=0.11 ~$M$_\odot$ and $M_{ej}=3.7~ $M$_\odot$. The window in the figure displays the evolution of the fraction of the $\gamma$-ray luminosity deposited into the envelope corresponding to the best fit. Of course the model fit does not account for the late drop in luminosity, however it reproduces well the initial part of the radioactive tail. SNe 1990I and 1993J, in this context, seem to have similar $^{56}$Ni ejected mass, whereas in view of the slight difference in brightness in favour of SN 1990I in the $[90:130]$ d time range, SN 1993J has a slightly smaller ejected mass ($\sim$ 3.4 M$_\odot$ from our best fit to SN 1993J data). The best fit parameters for SN 1993J are in good agreement with those from more complicated light curve calculations ($M(^{56}Ni)=0.1-0.14 ~$M$_\odot$ and $M_{ej}=1.9-3.5~ $M$_\odot$; Young, Baron $\&$ Branch. 1995\nocite{You95}) 

We note, however, that owing to the observed higher velocity nature of SN 1990I compared to the Ib SNe sample of Branch et al (2002), a higher kinetic energy might be anticipated though it is not, a priori mandatory. An adopted higher energy of about $E(10^{51} $ergs$)=1.2$ can be compensated with a mass $M_{ej}\simeq 3.95~ $M$_\odot$ to keep equal the $\gamma$-ray optical depth, in accordance with equation 3 (with an energy dependence given by: $M(^{56}Ni)\simeq0.11 \times $E$^{0.23} ~$M$_\odot$ and $M_{ej}\simeq 3.7 \times$E$^{0.36}~ $M$_\odot$).
Therefore, bearing in mind as well the possible uncertainties in the adopted parameters (i.e. distance and reddening), an ejected mass in the range $[3-4.5]$ M$_\odot$ is possible. 
  
\section{Discussion and conclusions}
We have presented and analysed photometric and spectroscopic data of SN 1990I, covering about 400 days of evolution. The observations were carried out at ESO and CTIO. The classification of the SN as a type Ib event discovered near maximum light is confirmed .

The optical He I lines are identified in the photospheric spectra of SN 1990I, displaying higher velocities compared to those derived, at similar epochs, from Branch et al. (2002) type Ib sample. The high velocity nature of SN 1990I is exemplified first from photospheric velocities, inferred from Fe II lines and He I lines, and second to a lesser extent from comparing [O I] 6300,64 \AA~emission line profiles with those of SNe 1993J and 1998bw.  The phase of
the spectral observations is uncertain because the first photometric point may
not define the point of maximum. But it cannot be pre-maximum. This means that
the spectra could only be around maximum and later in phase but not earlier. 
Thus the 
abnormally high expansion velocity  of SN 1990I seems beyond doubt
 {\it if} the phases for the other objects in the published sample are well determined. We
provided then an estimate of the maximum time and explosion date that are taken as
references throughout the paper. 

During the first 30 days the spectra at $6000-6500$ \AA~range show the presence of two absorption features near 6050 \AA~ and 6250 \AA~, while around 43 days since maximum light no presence of the 6250 \AA~ trough is seen. The possible identification of this trough with H$\alpha$ may point to the presence of a thin hydrogen layer. The presence of hydrogen, expelled at high velocities, in SN 1990I and in the ejecta of type Ib events in general is of course a crucial issue in understanding the progenitor properties. Deeper investigation of these line identifications requires detailed synthetic spectra and preferably a large sample study.

Entering the nebular phase, SN 1990I shows strong emission features of Mg I], Na I, [O I] and [Ca II], typical for type Ib/c SNe at late epochs. We note the presence of a depletion in the $[-2500:0]$ km s$^{-1}$ region of the [O I] 6300,64 \AA~emission profile in the 298 d and 358 d spectra, which may point to large scale clumps in the oxygen ejecta at late phases. On the other hand the possible presence of Fe II 6239 \AA~emission line feature (which seems common in many type Ib/c SNe at this phae) complicates the interpretation of the depletion feature.

The similarity in the main-sequence-mass between SN 1990I and a sample of SNe Ib/c is shown through the evolution of [Ca II] to [O I] flux ratio. It must be kept in mind, when comparing to type II SNe, that in Ib/c events there is no hydrogen rich [Ca II] emitting zone as is the case for type II objects and thus any firm conclusion should be taken with considerable caution.

We estimate a lower limit on the ejected oxygen mass to fall in the range $(0.7-1.35)$ M$_\odot$, although this estimate is subject to large uncertainties (i.e. density, temperature and fluxes). Clumping of the oxygen materials is indicated on the basis of two independent estimates of the density of oxygen. 

The large asymmetry seen, especially, in the oxygen and calcium forbidden lines is a result of radioactive material being distributed asymmetrically. Furthermore we identify possible fine structures superimposed on the top of some strong emission lines. In fact, contrary to SN 1993J where some bumps have been identified on the [O I] 6300,64 \AA~and O I 5577 \AA~profiles with similar velocities, two fine structures, on day 90, are found to repeat in the Ca II 8662 \AA~and [Ca II] 7307.5 \AA~ in SN 1990I. We interpret this to be an indication of different excitation conditions, in space and time, between SN 1990I and SN 1993J. 

Interestingly, SN 1990I spectra in addition to asymmetric profiles, display a blueshift change during the time period from 237 d to 258 d ($\Delta t/t \sim 0.1$), seen clearly in the [O I] 6300,64 \AA~ profile with a velocity of $\sim$600 km s$^{-1}$. Because of the presence of the depletion and poor S/N it is difficult to confirm the blueshift in the last two spectra. On the other hand, [Ca II] 7307.5 \AA~appears blueshifted at the last two epochs.

A photometric comparison of SN 1990I with other type Ib/c and IIb events, also demonstrates some peculiarities. During the early tail phase SN 1990I shows a decline rate similar to other SNe Ib/c, and faster than that of $^{56}$Co decay, indicating a low mass ejecta that allows the $\gamma$-rays to escape with lower deposition. Furthermore, light curves of SN 1990I show a marked change of slope. In fact, around day 230 the ``V, R and I'' light curves drop suddenly, while the B light curve tends to flatten. A comparison with events that show changes in slope is made introducing the third to second slope ratio as a measure of the how the SN changes its slope at late epochs. SN 1990I behaves in a completely different way compared to SN 1993J.

As a consequence the rapid and sudden drop is also seen in the colour evolution of SN 1990I, at an epoch similar to that of the already reported blueshift in the nebular lines. We interpret this rapid change, photometrically and spectroscopically, to be due to dust condensation in the ejecta of SN 1990I. The dust condensation at a time as early as $\sim$250 d (since explosion) could be due to the rapid cooling and low ejecta in SN 1990I compared to SN 1987A and SN 1999em which show evidence for dust formation around 530 d and 500 d, respectively. 

We constructed the ``$BVRI$'' quasi-bolometric light curve, and compared it to that of SN 1993J, revealing a similarity in the $[30:100]$ time interval. After 200 d, the two SNe evolve differently: while SN 1993J tends to flatten, SN 1990I falls dramatically. Then, in order to estimate physical parameters of the SN, we applied a simple $\gamma$-ray deposition model. The model allows us to estimate the ejecta and $^{56}$Ni masses ($M(^{56}Ni)=0.11 ~$M$_\odot$ and $M_{ej}=3.7~ $M$_\odot$). Still to note however that more accurate light curve modelling 
should include asymmetry and clumping effects.  

Drawing conclusions about the progenitor nature of SN 1990I (type Ib/c in general), based upon our 
present understanding of stellar evolution and the properties of the SN-explosion, is still speculative. 
The great unknown is whether this class of object has progenitor stars of similar (i.e $\sim 15-
 20~$M$_{\odot}$) or higher (i.e $> 20~$M$_{\odot}$) masses compared to type II SNe.

The oxygen mass derived for SN 1990I is of the order $\sim 1$ M$_{\odot}$, which is predicted to form in ZAMS stars
of $15-20~$M$_{\odot}$ according to the standard theory of evolution in massive stars. We note here that SN 1987A 
was found to eject $1.5-2~$M$_{\odot}$ of oxygen. This would mean that the SN 1990I progenitor was slightly lower in mass 
with respect to SN 1987A. Further support for the lower mass case comes from the estimated helium core 
($\sim 5\footnote{The ejected mass 3.7 M$_{\odot}$ and an assumed remnant of 1.3-1.6 M$_{\odot}$}~$M$_{\odot}$ for SN 1990I and $\sim 6~$M$_{\odot}$ for SN 1987A). What remains, however, is that SNe Ib/c eject
a mass of $^{56}$Ni in the range $0.1-0.15~$M$_{\odot}$, a higher amount compared to that from typical type II SNe
 (i.e. $0.07~$M$_{\odot}$ for SN 1987A). 

Although how the $^{56}$Ni mass scales with the progenitor mass remains 
unclear in core collapse SNe, a possible solution for this ``$ ^{56}Ni~ yield-M(progenitor)''$ paradox 
can be related to 
the binary nature of type Ib/c progenitors (Tutukov $\&$ Chugai. 1992\nocite{Tut92}) whereby the iron yield is very sensitive 
to rotation induced in the close binary system. In deed, very recently 
Hirschi et al. (2004)\nocite{Hir04} have tested the effects of rotation on evolution of massive stars.
The direct consequence of rotation on pre-SN models is o increase the core sizes especially
for stars in the $15-25~$M$_{\odot}$ mass interval. The $^{56}$Ni mass ejected in the explosion is found to increase by a factor of 1.5. 

It appears then that low mass progenitors within the context of close
binary system evolution (mass-loss due to mass transfer ) are more ``probable'' compared to single massive stars (i.e. Wolf-Rayet stars losing mass through winds). Clearly, large and well sampled type Ib/c SNe data (spectra and photometry) together with more precise estimates of M$_{ej}$, M($^{56}$Ni), M(oxygen) are needed in order to build an accurate physical understanding of SN Ib/c explosions.

The sudden fall in luminosity at day 230 results from the fact that the light curve in Fig. 11 is not a real bolometric light curve. After day 230 the dust absorbs at shorter wavelengths and re-emits at much longer IR wavelengths, consistent with the dust having an ambient temperature which is probably lower than 1500 K and decreasing with time. This contribution of IR radiation from the dust has not been included.
What remains to be understood, however, is why dust has formed in SN 1990I but has not been recorded in other type Ib/c. Possibly inadequate  sampling of light curves and spectra of type Ib/c at nebular phases is at least partially responsible. 

This paper demonstrates the need to obtain good temporal coverage in both photometry and spectroscopy with sufficient S/N to avoid compromising diagnostic capatibilities. 
\begin{acknowledgements}
This work is based on data obtained at CTIO and ESO-La Silla. We would like to 
thank all observational astronomers, in particular N. Suntzeff, M. Hamuy, M. Navarrete, H. Bond, A. Crotts and L. wells. A. Elmhamdi thanks B. Leibundgut
 for discussion and constructive comments on the original manuscript. A. Elmhamdi is grateful to the ESO (European Southern Observatory) support that allowed his stay at Garching, where this work was concluded. We are
 grateful to the referee (S. E. Woosley) for helpful comments.
\end{acknowledgements}

\end{document}